\documentstyle[11pt,epsf,amstex,righttag]{article}
\addtocounter{secnumdepth}{1}
\newcommand{\dd}{\text{d}}
\newcommand{\eps}{\varepsilon}
\newcommand{\p}{\partial}

\numberwithin{equation}{section}
\begin{document}

\thispagestyle{empty}
\title{Wilson renormalization\\ of a reaction--diffusion process} 
\author{F. van Wijland$^a$, K. Oerding$^b$, and H.J. Hilhorst$^a$\\\\
$^a$Laboratoire de Physique Th\'eorique et Hautes Energies$^1$\\
B\^atiment 211\\
Universit\'e de Paris-Sud\\
91405 Orsay cedex, France\\\\
$^b$Institut f\"ur Theoretische Physik III\\
Heinrich Heine Universit\"at\\
40225 D\"usseldorf, Germany\\}

\maketitle
\vspace{-1cm}
\begin{small}
\begin{abstract}
\noindent 
Healthy and sick individuals ($A$ and $B$ particles) diffuse
independently with diffusion constants $D_A$ and $D_B$.
Sick individuals upon encounter infect
healthy ones (at rate $k$), but may also spontaneously recover (at rate
$1/\tau$). The propagation of the 
epidemic therefore couples to the fluctuations in the total population
density. Global extinction 
occurs below a critical value $\rho_{c}$
of the spatially averaged total density. 
The epidemic evolves as 
the diffusion--reaction--decay process
$ A + B  \rightarrow   2B, \;\; B \rightarrow  A  $, for which we 
write down the field theory.  
The stationary state properties of this theory when $D_A=D_B$ were 
obtained by Kree {\it et al}.
The critical behavior for $D_A<D_B$ is governed by a new fixed point. 
We calculate the critical exponents of the stationary state
in an $\eps$ expansion, carried out by Wilson renormalization, 
below the critical dimension $d_{c}=4$. We then go on to
to obtain the critical initial time behavior at the extinction
threshold, both for
$D_A=D_B$ and $D_A<D_B$. 
There is nonuniversal dependence on the initial particle
distribution. The case $D_A>D_B$ remains unsolved.

\end{abstract}
\end{small}
\vspace{0.5cm}
\noindent L.P.T.H.E. - ORSAY 97/22\\
\\\\\\\\\\
{\small$^1$Laboratoire associ\'e au Centre National de la
Recherche Scientifique - URA D0063}
\newpage
\section{Introduction and phenomenological analysis}\label{Intro}
\subsection{Introduction}\label{Introduction}
We study the propagation of an epidemic in a population of fluctuating
density, with parameters set such that the epidemic is close to becoming
extinct.
The model is as follows.
Healthy and sick individuals, also called $A$ and $B$ particles, 
diffuse freely and independently on a lattice
of dimension $d$. Their diffusion constants $D_A$ and $D_B$ are in
general unequal. 
Sick individuals upon encounter infect healthy ones (at a rate $k$), but
may also spontaneously recover (at a rate
$1/\tau$).  The
epidemic therefore tends to propagate in regions of 
high total population density 
and to become locally extinct in regions of low density.
Below a critical density $\rho_{c}$
global extinction occurs, and an important question is how the
epidemic evolves near this threshold, which is analogous to a phase transition 
point in equilibrium critical phenomena. The role of the order parameter
is played by the $B$ particle density $\rho_B$. In particular, at threshold 
strong local fluctuations are expected, with a potentially
decisive impact on the global critical behavior.

The evolution of the epidemic may be formulated as the competition
between a reaction and a decay process of two diffusing chemical species,
\begin{equation}
A+B\stackrel{k}{\rightarrow}2B, \qquad B\stackrel{{1}/{\tau}}{\rightarrow} A
\label{AB}
\end{equation}
in which $A$ and $B$ stand for a healthy and a sick individual,
respectively. 

Let $m_i$ and $n_i$
be the number of $A$ and $B$ particles,
respectively, that occupy site $i$. We write $m=\{m_i\}$ and
$n=\{n_i\}$ for the two sets of occupation numbers and denote by
$P(m,n;t)$ the probability that at time $t$ the system is characterized
by exactly these sets. The rules of the process  
formulated above then define 
a master equation for the time evolution of $P(m,n;t)$ with fully
specified transition rates. The problem depends on the parameters $D_A,\,
D_B,\, k,\, 1/\tau$,\, and $\rho$. After scaling one may set $k=1$.\\

This problem 
is but one instance of a large class of such processes that all have the
following common features.
They are initially formulated in terms of 
a master equation which is hard, if not impossible, to solve.
The traditional approach is to write down rate equations for the
space and time dependent densities of the various species. This leads to 
numerous interesting systems of partial differential equations 
({\it reaction--diffusion equations}) 
that continue to be subjects of investigation.

The assumptions that lead to such 
reaction--diffusion equations can be justified only
in sufficiently high spatial dimension $d$. The reason is that
these equations constitute a {\it mean field}
description: the reaction terms  
involve the products of the {\it average} densities, rather than
the average of the product of the fluctuating densities. 

The general problem of statistical mechanics 
is therefore to determine the critical dimension
$d=d_c$ below which the mean-field approach becomes qualitatively incorrect,
and to describe the modifications that then appear.
Recent progress on such problems is due to the application of the
renormalization group (RG) to the formal
solution of the master equation, rewritten as a path integral and cast
in the form of a field theory. This will also be our procedure. 
It allows one, in an
expansion in the small parameter $d_c-d\equiv\eps$, to study the scale
invariant regimes of the problem and determine the associated exponents.

For the particular process of Eq.\,(\ref{AB}) 
the upper critical dimension appears to be $d_c=4$.
Questions of interest are first of all about
the properties of the stationary state and secondly about how the
system, prepared in a specified initial state, 
relaxes towards stationarity.
Of particular interest is the initial state with
the $A$ and $B$ particles distributed randomly 
with densities $\rho_A^{0}$ and $\rho_B^{0}$.
In that case, and in the limit of an infinite lattice, the $m_i$ and $n_i$ are
independent Poissonian random variables with 
averages $\rho_A^{0}$ and $\rho_B^{0}$,
respectively.\\ 

We now relate our work to existing literature.
We shall arrive at a field theory described by an action $S$ (see
Eq.\,(\ref{fullaction})) in which
\begin{equation}
\mu=\frac{D_A-D_B}{D_A}
\end{equation}
is a key parameter. The stationary state of a special case --
{\it viz.} with $D_A=D_B$ -- was 
studied recently by Kree, Schaub, and Schmittmann
\cite{KreeSchaubSchmittmann}.
These authors' physical interpretation is different: They start from 
a pair of Langevin equations
postulated for the evolution of a population density in a polluted
environment. The interest of both this work and of 
Ref.\,\cite{KreeSchaubSchmittmann} lies in the coupling between the
fluctuations of the total density and those of the density
of the species that is trying to survive (in our case
the $B$ particles).
If the fluctuations in the total density are suppressed,  
the full action $S$ becomes independent of $\mu$ and 
reduces to an action $S_{\text{DP}}$ known to represent both the
problem of directed percolation and 
the Schl\"ogl reaction. $S_{\text{DP}}$ has first appeared in particle physics
as Reggeon field theory.\\

This work goes beyond the existing results in the following ways.

({\it i}) The ratio $\mu$ appears to be relevant in
the sense of the RG. For $\mu<0$
the critical behavior is governed by a new fixed point. We determine the
new exponents to leading order in $\eps=4-d$. For $\mu>0$ the RG
equations do not have a fixed point
and we speculate about the interpretation of this fact.

({\it ii}) We consider relaxation to the stationary state at the
critical density $\rho=\rho_c$, first for the action $S_{\text{DP}}$ of
directed percolation and
then for the full action $S$ both with $\mu=0$ and $\mu<0$.
For asymptotically long times the order parameter $\rho_B(t)$ relaxes
towards zero as a power law of time whose exponent is a simple
combination ($\beta/\nu z$ in the usual notation) of the stationary
state exponents.
There is, however, an initial time regime -- which in
the limit $\rho_B^{0}\downarrow 0$ becomes
arbitrarily long -- where the order parameter {\it increases} as 
$t^{\theta'}$, where $\theta'$ is a new and independent {\it critical
initial slip} exponent. 
We calculate this exponent for an initial state in which the $m_i$ and
$n_i$ are independent but not
necessarily Poissonian random variables. We show that $\theta'$
is {\it nonuniversal}, as it depends continuously on the width
of the distribution.
The concept of critical initial slip was introduced by Janssen, Schaub,
and Schmittmann \cite{JanssenSchaubSchmittmann} (see also Janssen
\cite{Janssenreview}) 
and it is here for the first time, to our knowledge, that such an
exponent is calculated for a reaction--diffusion system.

({\it iii}) Finally, we innovate with respect to the existing literature
by not following the usual field-theoretic methods.
Instead, we perform
a Wilson renormalization, that is, we successively
integrate out shells of large momentum. We employ the dynamical version of
Wilson's method developed by Hohenberg and Halperin \cite{HohenbergHalperin}.
This method does not require any advanced 
knowledge of field theory. In the course of our work we check that in
known special cases earlier results are reproduced.

\subsection{Mean-field theory}
The mean field equations for the
densities $\rho_A$ and $\rho_B$ are
\begin{eqnarray}\label{meanfieldeqs}
\frac{\p\rho_A}{\p t}=D_A\Delta
\rho_A+\frac{1}{\tau}\rho_B-k\,\rho_A\rho_B\nonumber\\
\frac{\p\rho_B}{\p t}=D_B\Delta\rho_B-\frac{1}{\tau}\rho_B+k\,\rho_A\rho_B
\end{eqnarray}
By summing them one sees that the total density $\rho=\rho_A+\rho_B$ is
conserved. 
The spatially homogeneous solutions of these equations are easily found.
The total density $\rho$ appears to have
a threshold value $\rho=\rho_c^{\text{mf}}=(k\tau)^{-1}$ below which 
the stationary state consists of only $A$ particles. Above this threshold
density the all-$A$ state is still stationary but
unstable, and the stable stationary state has a nonzero
density of $B$ particles. Explicitly, Eq.\,(\ref{meanfieldeqs}) leads to 
\begin{equation}\begin{array}{llll}
\rho_A(\infty)=\rho_c^{\text{mf}},&\;&\rho_B(\infty)=\rho-
\rho_c^{\text{mf}}&\mbox{for }\rho>\rho_c^{\text{mf}}\\
\rho_A(\infty)=\rho,&\;&\rho_B(\infty)= 0&\mbox{for }\rho<\rho_c^{\text{mf}}
\end{array}\end{equation}
If we write, in analogy with critical phenomena in thermodynamic systems,
\begin{equation}
\rho_B(\infty)\stackrel{\rho\rightarrow  \rho_c}{\sim}(\rho-\rho_c)^{\beta}
\end{equation} 
then one sees that mean field theory gives $\beta=1$. 
The relaxation to the stationary state at criticality
from a homogeneous initial state with $B$-particle density
$\rho_B^{0}$ is easily derived from Eq.~(\ref{meanfieldeqs}),
\begin{equation}\label{rhoBtmf}
\rho_B(t)=\frac{\rho_B^{0}}{1+\rho_B^{0}kt}
\end{equation}
Again in analogy with critical relaxation
in thermodynamic systems we define
the exponents $z\nu$ and $\theta^{\prime}$ by
\begin{equation}
\rho_B(t)\sim t^{-\beta/(\nu z)}\mbox{ \quad for\, }t\gg (\rho_B^{0}k)^{-1}
\end{equation}
\begin{equation}
\rho_B(t)\sim \rho_B^{0}t^{\theta'} \mbox{ \quad for\, }\rho_B^{0}kt\ll 1
\end{equation}
Here $\theta'$ is the {\it critical initial slip} exponent introduced by
Janssen {\it et al.} \cite{JanssenSchaubSchmittmann,Janssenreview}.
Eq.\,(\ref{rhoBtmf}) shows that the mean field values are $z\nu=1$ and
$\theta'=0$. In the following sections, after showing that the
reaction--diffusion process of Eq.\,(\ref{AB}) has upper critical
dimension $d_c=4$, we will compute corrections
to the mean field values of these and other exponents to first order in 
$\eps = 4-d$.

\subsection{The master equation solved in terms of a path integral}
In order to go beyond mean field theory,
one has to start from the master equation defined implicitly in
subsection \ref{Introduction}.
It is convenient to introduce an 
orthonormal basis of states $|m,n\rangle$ and work in the Hilbert space
generated by this basis. The 
probability distribution $P(m,n;t)$ is then associated with the state
\begin{equation}
|P(t)\rangle=\sum_{m,n}P(m,n;t)|m,n\rangle
\end{equation}
For an appropriately constructed time evolution operator $\hat{H}$
the master equation for $P(m,n;t)$ is equivalent
to the imaginary time Schr\"odinger equation
\begin{equation}
\frac{\text{d}}{\text{d}t}|P(t)\rangle=-\hat{H}|P(t)\rangle
\label{evoleqn}
\end{equation}
From the formal solution of Eq.~(\ref{evoleqn}),
\begin{equation}\label{evolsoln}
|P(t)\rangle=\text{e}^{-\hat{H}t}|P(0)\rangle
\end{equation}
all properties of the reaction--diffusion system may be derived.
Eq.\,(\ref{evolsoln}) determines in particular in the limit 
$t\rightarrow\infty$ the stationary state.  
The combined efforts of many workers, {\it e.g.}
\cite{Doi,Peliti,LeeCardy}, have led to 
methods of converting the formal solution of type (\ref{evolsoln}) 
into expressions that can be analyzed. 
We shall use here one such method, which has now become common. 
The first step is to
express $\hat{H}$ in terms of
creation and annihilation operators for the two particle species:
$a_i^{\dag}$ and $a_i$ for the $A$ particles, and $b_i^{\dag}$ and $b_i$
for the $B$ particles. These operators can be
defined such that they have boson commutation relations.
Next, the exponential in (\ref{evolsoln}) is subjected to time slicing
with the aid of a coherent state representation of the 
harmonic oscillator operators.
As a result, and in the limit where the site index becomes a continuous
space vector $x$, there appear  
space and time dependent classical fields $a^\ast(x,t),
\;a(x,t)$ and $b^\ast(x,t), \;b(x,t)$ that are 
directly associated with the two particle types.
The solution (\ref{evolsoln}) finally takes the form of a 
path integral on these fields, weighted with the exponential of an
action $S$:
\begin{equation}
|P(t)\rangle=\int{\cal D}a {\cal D}a^\ast {\cal D}b {\cal D}b^\ast
\;\text{e}^{-S[a,a^\ast,b,b^\ast]}|P(0)\rangle
\end{equation}
The exact way in which such physical quantities as local densities, 
correlation and response functions can be obtained from the fields  is
detailed in \cite{LeeCardy}.
\subsection{Action and scaling dimensionalities}
The full action for the problem of Eq.\,(\ref{AB}) is
\begin{equation}\label{actiondedepart}
\begin{split}
S[a,a^\ast,b,b^\ast]=&\int d^dx\,
dt\Big[a^\ast(\p_t-D_A\Delta)a+b^\ast(\p_t-D_B\Delta)b\\&+kab(a^\ast-
b^\ast)b^\ast-\frac{1}{\tau}(a^\ast-b^\ast)b\Big]\\&-\int
d^dx\Big[\rho_A^{0}a^\ast(x,0)+\rho_B^{0} b^\ast(x,0)+a(x,T)+b(x,T)\Big]
\end{split}
\end{equation}
Here $\rho_A^{0}$ and $\rho_B^{0}$ are the initial $A$ and $B$
particle densities,
and Poissonian distribution of the
initial occupation numbers has been assumed; and the time
integral runs through a fixed interval $[0,T]$. 
The averages with weight $\text{e}^{-S}$ of the fields $a(x,t)$
and $b(x,t)$ are equal \cite{Doi,Peliti,LeeCardy} to
the local densities of the $A$ and $B$ particles. In order
to cast Eq.~(\ref{actiondedepart}) in a form suitable to
subsequent analysis, we first change from $a^*$ and $b^*$ to the 
variables
\begin{equation}
\begin{array}{rclcrcl}
\overline{a}&=&a^\ast-1&\text{and}&
\overline{b}&=&b^\ast-1
\end{array}
\end{equation}
the effect of which is to get rid of the boundary terms at time $t=T$.
Then we introduce the fields 
\begin{equation}
\begin{array}{rclcrcl}
\varphi&=&a+b-\rho\Theta(t)&&
\overline{\varphi}&=&\overline{a}\\
\psi&=&b&&
\overline{\psi}&=&\overline{b}-\overline{a}
\end{array}
\end{equation}
where the parameter $\rho$ denotes as before the average total particle
density, and $\Theta(t)$ is the unit step function. After further rescaling
of the fields and time one finds the expression for the action that will
be the starting point in the following sections,
\begin{equation}\label{fullaction}
\begin{split}
S{[}\varphi,\overline{\varphi},\psi,\overline{\psi}{]}=&\int d^dx dt\Big[
\overline{\varphi}(\p_t-\Delta)\varphi+\overline{\psi}(\p_t+
\lambda(\sigma-\Delta))\psi\\&+\mu\overline{\varphi}\Delta\psi+g\psi
\overline{\psi}(\psi-\overline{\psi})+u\psi\overline{\psi}(\varphi+
\overline{\varphi})\\&+v_1\big(\psi\overline{\psi}\big)^2+v_2\psi
\overline{\psi}(\psi\overline{\varphi}-\overline{\psi}\varphi)+
v_3\varphi\overline{\varphi}\psi\overline{\psi}\\
&-\rho_B^{0}\delta(t)\;\overline{\psi}\Big]
\end{split}
\end{equation}
In terms of the original parameters of the master equation the coupling 
constants of this action are given by
\begin{equation}\begin{array}{llll}
\mu=1-D_B/D_A&g=k\sqrt{\rho}/D_A&u=-k\sqrt{\rho}/D_A&\\
v_1=v_2=-v_3=k/D_A&\lambda\sigma=k(\rho_c^{\text{mf}}-\rho)/D_A&\lambda=
{D_B}/{D_A}\\\rho_B^{(0)}=\rho_B^{0}/\sqrt{\rho}&&&
\end{array}\end{equation}
If one omits from the action Eq.~(\ref{fullaction}) the initial time
term proportional to $\rho_B^{(0)}$,
then the remainder is, for $\mu=0$, invariant under the time reversal symmetry
\begin{equation}\label{sym}
\begin{array}{rcl}
\psi(x,t)&\rightarrow&-\overline{\psi}(x,-t)\\
\overline{\psi}(x,t)&\rightarrow&-\psi(x,-t) \\
\varphi(x,t)&\rightarrow&\phantom{-}\overline{\varphi}(x,-t)\\
\overline{\varphi}(x,t)&\rightarrow&\phantom{-}\varphi(x,-t)
\end{array}
\end{equation}
The breaking of this symmetry for
$\mu\neq 0$, that is, when the diffusion constants $D_A$ and $D_B$
are different, will be seen to have 
far reaching consequences for the critical behavior
of this system.
The naive scaling dimensionalities of the coupling constants are, in
powers of an inverse length,
\begin{equation}
[g]=[u]=2-d/2,\quad [v_i]=2-d,\quad [\mu]=0
\label{naivescaling}
\end{equation}
This shows that $d_c=4$ is the upper critical dimension for this problem
and that in an expansion around $d=4$ we may drop the $v_i$
vertices.\\

\noindent The terms occurring in Eq.~(\ref{fullaction}) may be given the
following 
physical interpretation. The $g$ vertex, which involves only the $B$ 
fields $\psi$ and $\overline{\psi}$, accounts for fluctuations in the
contamination process ~~$ A+B\rightarrow2B $~~
due to fluctuations in the $B$ particle density, and as though the $A$
particles formed a homogeneous background. The sum 
$\varphi+\overline{\varphi}$ may be shown to be equal to the fluctuating
part of the total density. 
The $u$  vertex couples  this to the density of $B$ particles,
expressing that the reaction ~~$ A+B\rightarrow 2B $~~ is also affected
by fluctuations of the $A$ particle density. \\

\subsection{A reduced problem: Directed Percolation}

If in the action $S$ of Eq.~(\ref{fullaction}) we replace the fields
$\varphi$ and $\overline{\varphi}$ by their mean field values $\varphi=0$ and
$\overline{\varphi}=0$, and drop the fourth order terms that were argued
above to be irrelevant, the result is the action $S_{\text{DP}}$ known
in particle physics as the {\it Reggeon field theory} and given by
\begin{equation}
S_{\text{DP}}[\psi,\overline{\psi}]=\int d^dx\,
dt\Big[\overline{\psi}(\p_t+\lambda(\sigma-\Delta))\psi+g\psi
\overline{\psi}(\psi-\overline{\psi})-\rho_B^{(0)}\;\delta(t)\overline{\psi}
\Big]
\label{actionDP}
\end{equation}
Physically this means that we have constrained the total density to be
constant (equal to $\rho$) in space and time. If the $B$ density is much
lower than the $A$ density, as is the case near criticality, this
constraint means that the $B$ particles interact with a continuous
background. The first link between the action $S_{\text{DP}}$ and statistical 
mechanics was established by
Cardy and Sugar \cite{CardySugar}, 
who showed that it describes the
problem of directed bond percolation. 
Subsequently it was shown by Grassberger and Sundermeyer
\cite{GrassbergerSundermeyer} and by Grassberger and de la Torre
\cite{GrassbergerdelaTorre}  (see also Janssen \cite{J81})
that this action also describes Schl\"ogl's
chemical reaction \cite{Schlogl}.
The $\eps$ expansion for the exponents of the action $S_{\text{DP}}$
was known from field theory (\cite{Bronzan} and references therein)
and has been extended by Janssen \cite{J81}.

The remainder of this article is organized as follows. In Section
\ref{DP}, for a twofold reason, we return to the action
$S_{\text{DP}}$.  
We consider for the first time its relaxation from an initial
state and calculate the
initial critical slip exponent $\theta'$.
Secondly, since throughout this work we employ time dependent Wilson
renormalization, we explain our method on the example of the relatively
simple action $S_{\text{DP}}$, after which we apply it with only few
extra comments to the full action $S$ of Eq.\,(\ref{fullaction}).

In section \ref{Fullproblem} we return to the full action $S$.
We determine its stationary state
properties, which are then used to study the relaxation from the
initial state. 
We also study the effect of
non-Poissonian initial particle distributions.
\section{Directed Percolation: Wilson renormalization and
critical initial slip exponent}\label{DP}
The stationary state of the action
Eq.~(\ref{actionDP}) exhibits critical behavior that has been studied 
by several authors both within the framework of an
$\eps$-expansion around its upper critical dimension $d=4$.
The analysis of the action Eq.~(\ref{actionDP}) is usually performed ({\it
e.g. }\cite{CardySugar,J81}) by
means of field-theoretic methods, based upon the
perturbative expansion of vertex functions and dimensional
regularization, and leading to Callan-Symanzik equations. In this work,
instead, we apply Wilson's dynamic renormalization group,
that is, successively eliminate
short wavelength degrees of freedom. This
method has already been used successfully to study the dynamics of spin
systems for instance; it is applied here for the first time to a
reaction--diffusion problem. 

\subsection{Wilson's dynamic renormalization group}\label{WilsonRG}
The extension of Wilson's renormalization procedure to time dependent
systems has been
described by Hohenberg and Halperin \cite{HohenbergHalperin}. 
We recall here how this method works on the example of the action
$S_{\text{DP}}$ of Eq.\,(\ref{actionDP}). 
For a study restricted to the stationary state the boundary term at $t=0$ 
may be suppressed. The action may then be written as
$S_{\text{DP}}=S_0+S_{\text{int}}$ with the free term $S_0$ given by
\begin{equation}
S_0[\psi,\overline{\psi}]=\int\overline{\psi}(\p_t+\lambda(\sigma-\Delta))\psi
\end{equation}
The free propagator, in terms of the spatial Fourier components, is 
\begin{equation}
\langle\overline{\psi}(-k,t')\psi(k,t)\rangle=\Theta(t-t')\text{e}^{-(k^
2+\sigma)(t-t')}
\end{equation}
\noindent One is interested in averages carried out with the
weight $\text{e}^{-S_{\text{DP}}}$. In Wilson's procedure this
weight is first integrated over the Fourier components of
the fields with wavevectors $k$ in a momentum shell $\Omega_\Lambda$
defined by 
\begin{equation}
\Omega_\Lambda=\{k\,|\,\Lambda/b<k<\Lambda\}
\end{equation}
We denote these fields for brevity by $\psi_>$ and $\overline{\psi}_>$
and the remaining fields by $\psi_<$ and $\overline{\psi}_<$. 
The scale factor $b$ will be taken infinitesimally close to 1,
and the momentum space cut-off $\Lambda$ is of order unity.  One obtains
an effective action $S_{\text{DP}}^{\text{eff}}$
\begin{equation}
{\text{e}}^{-S_{\text{DP}}^{\text{eff}}[\psi_<,\overline{\psi}_<]}=\int{\cal
D}[\psi_>,\overline{\psi}_>]{\text{e}}^{-S_{\text{DP}}}
\end{equation}
In practice the integration step is carried out 
perturbatively in $S_{\text{int}}$ by using
the cumulant expansion
\begin{equation}
\label{cumexp}
\exp{(-S_{\text{DP}}^{\text{eff}})}=\exp{(-\langle
S_{\text{int}}\rangle^\prime+\frac{1}{2!}\langle S_{ 
\text{int}}^2\rangle_c^\prime-\frac{1}{3!}\langle
S_{\text{int}}^3\rangle_c^{\prime}+\ldots)}
\end{equation}
The index $c$ indicates a cumulant and
$\langle\ldots\rangle'$ 
stands for an average on the $>$ fields with
weight ${\text{e}}^{-S_0}$. Therefore there
remain only Gaussian integrations to perform.

Wick's theorem ensures that the integrations reduce to
pair contractions,
and the graphical rules for dealing with these are
easily found. 
It is convenient (see Fig.\,1) to represent a
$\overline{\psi}$ field (a $\psi$ field) by a leg with (without) an arrow.
The interaction term $S_{\text{int}}$ is a sum of
three-leg vertices of types $g_1$ and $g_2$ in Fig.\,1. The $n$th order
cumulant in Eq.~(\ref{cumexp}) 
comes from diagrams with $n$ such vertices.
An external leg of a diagram will stand for a
$<$ field and an internal line for the contraction of two $>$ fields. 
Each vertex is integrated
over space and time. Only
one-particle irreducible diagrams need to be considered.
Moreover, any diagram in which one can go around a loop in the 
direction of the arrows vanishes by causality. 
A contraction of any pair of fields at equal times is zero, 
a property conveniently expressed by the convention $\Theta(0)=0$ and
commented upon {\it e.g.} in Ref.\,\cite{Janssenreview}. Finally, 
products of noncontracted fields at different times are 
reduced to single-time expressions by Taylor expansion in the time
difference.\\

\noindent After the integration on the momentum shell fields is done one
applies to $S_{\text{DP}}^{\text{eff}}$ a scale
transformation which depends on exponents $z,\, d_\psi,\,$ and\,
$d_{\overline{\psi}}\,$ still to be determined, 
\begin{equation}\begin{array}{ll}
x=bx'&\qquad   \psi_<(x,t)=b^{-d_\psi}\psi'(x',t')\\
t=b^zt'&\qquad  \overline{\psi}_<(x,t)=b^{-d_{\overline
{\psi}}}\,\overline{\psi}\,'(x',t') 
\end{array}\end{equation}
and drops the primes of the new variables. The result is
a renormalized action whose coupling constants are given in terms of the
original couplings. 
In dimension $d=d_c-\eps$ the cumulant series in Eq.\,(\ref{cumexp}) 
usually 
corresponds to an expansion in powers of $\eps$ and upon truncating it at
the desired order one has a RG involving only a finite number of
coupling constants.
If a fixed point of the transformation can be found, then 
scale invariance is a direct consequence and scaling laws follow. 

\subsection{Stationary state of $S_{{\footnotesize\mbox{DP}}}$}
The time reversal symmetry of
Eq.~(\ref{sym}) imposes that $d_\psi=d_{\overline{\psi}}$.
We write $d_\psi=d_{\overline{\psi}}=\frac{1}{2}(d+\eta)$, where $\eta$ is
the anomalous dimension of the fields $\psi$ and
$\overline{\psi}$. We have applied the renormalization 
procedure of subsection \ref{WilsonRG} to $S_{\text{DP}}$, taking into
account all diagrams with two and three vertices. 
The resulting 
recursion relations read, with $b\equiv{\text{e}}^\ell$, 
\begin{equation}\label{DPdg}
\frac{\dd g}{\dd\ell}=g\Big[z-2+\frac{\eps}{2}-\frac{3\eta}{2}-2g^2\frac{K_4}
{\lambda^2}\Big]
\end{equation}
for the coupling constant of the three-leg vertex, and
\begin{equation}
\frac{\dd\sigma}{\dd\ell}=(z-\eta)\sigma+\frac{g^2 K_4}{\lambda}\frac{\Lambda^4}
{\Lambda^2+\sigma}
\end{equation}
for the deviation from the critical density. Here $K_4$ is the surface
area of the unit sphere in $\Bbb{R}^4$ divided by $(2\pi)^4$.
Requiring that the coefficients of the terms with 
the time and space derivatives in
$S_0$ remain constant leads to the conditions
\begin{eqnarray}\label{dtfixed}
-\eta-g^2\frac{K_4}{2\lambda^2}=0
\\
\label{Deltaxfixed}
z-2-\eta-g^2\frac{K_4}{4\lambda^2}=0
\end{eqnarray}
These allow one to express 
the exponents $z$ and $\eta$ in terms of $g^2\lambda^{-2}$.
Eqs.~(\ref{DPdg}-\ref{Deltaxfixed}) have a nontrivial fixed point, 
\begin{equation}\label{gDPfp}
g^{\star2}=\frac{\lambda^2}{3K_4}\;\eps
\end{equation}
where we neglect terms of higher order in $\eps$.
This fixed point is stable since in its vicinity $g$ scales with the
negative exponent
$y_g=-\eps$. The fixed point value of $\sigma$ is
\begin{equation}\label{212}
\sigma^\star=-\frac{\lambda\Lambda^2}{6}\;\eps
\end{equation}
In view of the definition of $\sigma$ this implies that the critical
density $\rho_c$ in dimension $4-\eps$ is higher than the mean field
value $\rho_c^{\text{mf}}=1/k\tau$.
Near the fixed point $\sigma$ and time scale with the exponents 
\begin{equation}
y_\sigma=\nu^{-1}=2-\frac{\eps}{4}
\label{ysigma}
\end{equation}
\begin{equation}
z=2-\frac{\eps}{12}
\end{equation}
respectively, and the anomalous dimension of the fields $\psi$ and
$\overline{\psi}$ is  
\begin{equation}
\eta=-\frac{\eps}{6}
\label{reseta}
\end{equation}
In Eq.\,(\ref{ysigma}) $\nu$ is the usual correlation length exponent.
Since $\beta=\frac{1}{2}\nu(d+\eta)$, the density of the $B$ species
scales with
\begin{equation}
\beta=1-\frac{\eps}{6}
\end{equation}
The analysis has been extended to second order in $\eps$ by
Janssen~\cite{J81}.\\

\noindent The logarithmic correction
in dimension $d=4$,
\begin{equation}
\rho_B(\infty)\sim\sigma\ln^{\frac{1}{3}}\sigma
\end{equation}
has not, to our knowledge, been mentioned before.

\subsection{Relaxation to the stationary state of Directed
Percolation}\label{relaxDP}
Relaxation to the stationary state of the directed percolation problem
has not been considered before but can be studied by the same
techniques.
One has to keep the initial time term, with "coupling constant"
$\rho_B^{(0)}$, in the action of Eq.\,(\ref{actionDP}).
This term represents an initial state with Poissonian distribution of
the $B$ particles.
In the diagrams it will be shown as an arrowed leg starting from a
$t=0$ vertex indicated by an empty circle (see Fig.\,1). 
It is easy to extend the analysis to a non-Poissonian initial distribution
of particles. In that case we have to include in the action $S_{\text{
DP}}$ of Eq.\,(\ref{actionDP}) another initial time term, {\it viz.}
\begin{equation}
\Delta_\psi\,\int\,d^dx\;\overline{\psi}^2(x,0)
\label{Dpsiterm}
\end{equation}
Higher order terms in the field $\overline{\psi}$ also appear, but are
irrelevant under renormalization.
The coupling constant $\Delta_\psi$ in Eq.\,(\ref{Dpsiterm}) is related
to the width of the distribution by
\begin{equation}
\Delta_\psi=-\frac{1}{2\rho}[\langle\Delta n_i^2\rangle - \langle n_i\rangle]
\label{Deltapsi}
\end{equation}
which vanishes for a Poisson distribution. 
The naive scaling dimension of the $\Delta_\psi$ term is easily found to
be zero. This means
that it is marginal and has to be included
in the RG calculation.
Any other terms generated by these two initial terms are 
irrelevant. 
Fig.\,2 shows the two diagrams that yield renormalization contributions
to $\rho_B^{(0)}$ and to $\Delta_\psi$ which are {\it linear} in these
quantities. The conclusions of this section may be arrived at on the basis
of only these
diagrams. Nevertheless we shall complete the calculation to one-loop
order by also computing the contributions of the
diagrams of Fig.\,3.
The diagram of Fig.\,3(a) is the most complicated one to consider and
the only one that we shall discuss explicitly. Its expression is
\begin{equation}
2\Delta_\psi\frac{g^2K_4}{\lambda^2}\!\!\int\!\!\frac{d^dk}{(2\pi)^d}\!
\int\!dt\;d\tau\;\psi(k,t)\overline{\psi}(-k,t+\tau)\!\int_{\Omega_{\Lambda}}
\!\!\frac{d^dq}{(2\pi)^d}\text{e}^{-2\lambda q^2
t-\lambda(q^2+(k-q)^2+\sigma) \tau} 
\end{equation} 
One expands the fields in the integral for small $t$ and $\tau$, and
the exponential in powers of $k$. One can then carry the time
integrations out explicitly and use the initial
time constraint \cite{JanssenSchaubSchmittmann}
\begin{equation}
\psi(k,t=0)=\rho_B^{(0)}-2\Delta_\psi\overline{\psi}(k,t=0)
\end{equation} It appears that out of all terms only 
\begin{equation}
2\Delta_\psi\frac{g^2K_4}{\lambda^2}\rho_B^{(0)}\int d^dx 
\;\overline{\psi}(x,0)
\end{equation}
renormalizes $\rho_B^{(0)}$ and 
\begin{equation}
-4\Delta_\psi^2\frac{g^2K_4}{\lambda^2}\int d^dx \;\overline{\psi}^2(x,0)
\end{equation}
renormalizes $\Delta_\psi$. The other terms are
irrelevant. The recursion
relations for the couplings $\rho_B^{(0)}$ and $\Delta_\psi$ that one
finally finds are, to one loop order,
\begin{eqnarray}
\frac{\dd\rho_B^{(0)}}{\dd\ell}=\frac{1}{2}(d-\eta)\rho_B^{(0)}+2\Delta_\psi
 g \frac{K_4
\Lambda^2}{2\lambda(\Lambda^2+\sigma)}-2\rho_B^{(0)}\Delta_\psi\frac{g^2
K_4}{\lambda^2}
\label{RGrhoB0}
\\
\frac{\dd\Delta_\psi}{\dd\ell}=\Delta_\psi(-\eta-2g^2\frac{K_4}{\lambda^
2}-4g^2\Delta_\psi\frac{K_4}{\lambda^2}- \Delta_\psi g^2\frac{K_4}{\lambda^2})
\label{RGDeltapsi}
\end{eqnarray}
The first term in Eq.\,(\ref{RGrhoB0}) and the term proportional to
$\eta$ in Eq.~(\ref{RGDeltapsi}) are due to rescaling of the fields and
the variables. 
The remaining terms in Eq.\,(\ref{RGrhoB0}) come from the diagrams of Figs.\,2a
and 3a, and those 
in Eq.~(\ref{RGDeltapsi}) from the diagrams of Figs.\,2b, 3a, and 3b,
respectively. 
Upon combining Eq.\,(\ref{RGDeltapsi}) with Eqs.\,(\ref{gDPfp}) and
(\ref{reseta})
one sees that only the fixed point $\Delta_\psi^\star=0$  is stable,
and $\Delta_\psi$ scales with the negative exponent 
$y_{\Delta_\psi}=-\frac{\eps}{2}$ and hence is 
irrelevant. 
Substitution of this fixed point value in Eq.\,(\ref{RGrhoB0}) 
shows that $\rho_B^{(0)}$ scales with the exponent
\begin{equation}
y_B=\frac{1}{2}(d-\eta)
\end{equation}
We note in passing that, since the same diagrams renormalize
$g\psi\overline{\psi}^2$ and $\Delta_\psi\overline{\psi}^2\big|_{t=0}$, 
comparison of Eqs.\,(\ref{RGDeltapsi}) and (\ref{DPdg}) shows that the identity
\begin{equation}
y_{\Delta_\psi}=2-z-\frac{\eps}{2}+\frac{\eta}{2} \label{221}
\end{equation}
must hold to every order in $\eps$.

\subsection{Initial and long time scaling of the relaxation}\label{relaxDP2}
We analyze the relaxation towards the stationary state
following Diehl and Ritschel 
\cite{DiehlRitschel} (see also \cite{RitschelDiehl}). To simplify the
discussion we place ourselves at the critical density, that is, put
$\sigma=0$. Since $\rho_B(t)$ is
equal to the average of $\psi(x,t)$ with respect to ${\text
e}^{-S_{\text{DP}}}$, it scales with the exponent $(d+\eta)/2$ and 
one has the scaling behavior 
\begin{equation}
\rho_B(t)=b^{-\frac{d+\eta}{2}}F(b^{-z}t,b^{\frac{d-\eta}{2}}\rho_B^{(0)})
\end{equation}
valid in the limit $b\rightarrow\infty$ and for fixed arguments of the
function $F$. Choosing $b^z=t$ and employing the relation
$\beta=\nu(d+\eta)/2$ leads to 
\begin{equation}
\rho_B(t)=t^{-\frac{\beta}{\nu z}} F(1,\rho_B^{(0)}t^{\frac{\beta}{\nu
z}+\theta'}) 
\end{equation}
In the limit of large time $t$, at fixed initial density $\rho_B^{(0)}$,
one obtains the usual critical relaxation proportional to $t^{\beta/\nu
z}$ at the condition that $F(x,y)\sim 1$ for $y\rightarrow\infty$.
In the limit of small $\rho_B^{(0)}$ at fixed time $t$ 
a scaling regime appears whose existence was first pointed out by
Janssen {\it et al.} \cite{JanssenSchaubSchmittmann}. In this limit
the density
$\rho_B(t)$ should be proportional to $\rho_B^{(0)}$, which is possible
only if $F(x,y)\sim y$ in the small $y$ limit. Hence we can write
\begin{equation}
\rho_B(t)=\rho_B^{(0)}t^{\theta'}
{\cal F}(\rho_B^{(0)}t^{\frac{\beta}{\nu z}+\theta'})
\end{equation}
in which ${\cal F}(y)=y^{-1}F(1,y)$ and
\begin{equation}
\theta'=-\frac{\eta}{z}=\frac{\eps}{12},
\label{thetaprime}
\end{equation}
where the function ${\cal F}$ has the limit behavior
\begin{equation}
{\cal F}(y)\sim 1 \quad (y\rightarrow 0), \qquad {\cal F}(y)\sim y^{-1}
\quad (y\rightarrow\infty) 
\end{equation}
Here $\theta'$ is the critical initial slip exponent
\cite{JanssenSchaubSchmittmann}. The fact that it is positive here
implies that in an initial time regime (excluding microscopically small
times) the $B$ particle density {\it increases} as
\begin{equation}
\rho_B(t)\sim \rho_B^{(0)}t^{\theta'}
\end{equation}
Crossover to the regime of asymptotically long times takes place at
$t\sim \tau_{\text{cross}}$, with
\begin{equation}
\tau_{\text{cross}}\sim(\rho_B^{(0)})^{-1/(\frac{\beta}{\nu z}+\theta')}
\end{equation}
The positivity of $\theta'$ and the initial rise of $\rho_B(t)$ 
have been discussed and interpreted in
Refs.\,\cite{JanssenSchaubSchmittmann,DiehlRitschel}. They can be qualitatively
understood by the lack of critical fluctuations in the initial state.
In the case of the reaction--diffusion systems considered here the
fluctuations in the stationary state are responsible for the shift of
the critical density to a larger value compared its mean field
value. This is reflected by the negative $\sigma^{\star}$ obtained in
Eq.~(\ref{212}). Since the density
distribution in the initial state is Poissonian and thus free of long
range correlations the system shows qualitatively a mean field-like
behaviour during the initial stage of the relaxation.
For $\sigma = \sigma^{\star}<0$ the density of B-particles therefore increases
at (macroscopically) short times~$t$. Notice, however, that this
qualitative argument applies generally not to
{\em microscopically} short times~\cite{RC}.

An important point, stressed in Ref.\,\cite{JanssenSchaubSchmittmann} 
and not apparent from the preceding discussion, is that
$\theta'$ is an {\it independent} critical exponent. Whereas in the special
case of $S_{\text{DP}}$ considered here it could be expressed as
$-\eta/z$ (see Eq.\,(\ref{thetaprime})), that relation does not hold in
general. This will be seen, in particular, in Section \ref{Fullproblem}.

\noindent The scaling function ${\cal F}$ is known at the mean-field level, 
\begin{equation}
{\cal F}(y)=\frac{1}{1+\text{constant}\times y}
\end{equation}
In dimension $d=4$ one finds that the $B$ density decays as
\begin{eqnarray}
\rho_B(t)&\sim&\rho_B^{(0)}\;\ln^{\frac{1}{6}}t \quad
(t\ll\tau_{\text{cross}})\nonumber\\ 
\rho_B(t)&\sim& t^{-1}\ln^{\frac{1}{2}}t \quad (t\gg\tau_{\text{cross}}) 
\end{eqnarray} 
where $\tau_{\text{cross}}\sim
(\rho_B^{(0)})^{-1}\ln^{\frac{1}{3}}(1/\rho_B^{(0)})$. 

Finally, when the system is slightly off criticality, that is, for $\sigma\neq
0$, the scaling function ${\cal F}$ depends on the supplementary scaling
variable $|\sigma|t^{1/\nu z}$. The power law decay $\sim t^{-\beta/\nu
z}$ is then cut off exponentially on a time scale of order $\xi^z$,
where $\xi\sim \sigma^{-\nu}$ is the correlation length.

\section{Wilson renormalization of the full action $S$}\label{Fullproblem}
\subsection{Introduction}
The RG procedure applied above to the action $S_{\text{DP}}$ of the
Directed Percolation problem can be extended without difficulty to the
action $S$ given by Eq.\,(\ref{fullaction}) and describing 
the full problem defined in Section \ref{Introduction}.
This action contains a parameter $\mu$ which distinguishes between the
diffusion constants of the healthy and the sick individuals.
It appears to constitute an immediate and natural step beyond the action
of Directed Percolation:
Led by an interpretation different from ours, and starting from a pair
of Langevin equations postulated for a problem in population dynamics,
Kree, Schaub and Schmittmann
\cite{KreeSchaubSchmittmann} arrive at the same action but specialized
to $\mu=0$. Using field-theoretic techniques they analyze its stationary
state and compute  
the critical exponents $z$, $\eta$, and $\nu$ to order $\eps$. 

Below we first derive the
stationary state RG equations for the general case $\mu\neq 0$. 
Since $\mu$ appears to be relevant, a new analysis is needed. For
$\mu<0$ we find a new fixed point and calculate the new critical
exponents to lowest order in $\eps$. 

Next we derive the RG equations for the additional couplings that
describe the initial state. We shall, in particular, consider an initial
particle distribution which is independent from site to site, but
non-Poissonian.\\ 

\subsection{Stationary state of the full action $S$}

Our starting point is the action $S$ of Eq.\,(\ref{fullaction}). 
Since our first interest is in the stationary state, we ignore for the
moment the initial time term proportional to $\rho_B^{(0)}$.
As before we write $S=S_0+S_{\text{int}}$. The free part $S_0$ of the
action now includes the term in Eq.\,(\ref{fullaction}) that is proportional to
$\mu$, and whose presence is due to the diffusion constants of the $A$
and $B$ particles being unequal. This term violates the time reversal
symmetry of Eq.\,(\ref{sym}) and introduces a nonzero correlator
\begin{equation}
\langle\varphi(k,t_1)\overline{\psi}(-k,t_2)\rangle=\frac{\mu
k^2}{(\lambda-1)k^2+\lambda\sigma}\Theta(t_1-t_2)({\text
e}^{-k^2(t_1-t_2)}-{\text e}^{-\lambda(k^2+\sigma)(t_1-t_2)})
\end{equation}
Diagrammatically this contraction is represented by a solid leg with an
arrow that connects to a dashed leg without arrow to form an internal
line (see Fig.\,1).
As a consequence the symmetry between the $\psi^2\overline{\psi}$ and
$\psi\overline{\psi}^2$ vertices in the action is broken and we 
write the interaction part $S_{\text{int}}$ as 
\begin{equation}
S_{\text{int}}=\int\Big[g_1\psi^2\overline{\psi}-g_2\psi\overline{\psi}^
2+u\psi\overline{\psi}(\varphi+\overline{\varphi})\Big]
\end{equation}
with independent $g_1$ and $g_2$.

Some further general considerations are useful.
The vertices of $S_{\text{int}}$ (see Fig.\,1) do not 
allow for renormalization of the
$\varphi$ propagator. We therefore conclude that
\begin{equation}
z=2
\end{equation}
In order to show that this is the dynamic scaling exponent not only 
of the conserved density $\varphi$ but also of the fluctuations of the
order parameter, 
one has to check that the coupling $\lambda$ tends to finite nonzero 
fixed point value (see below).
From dimensional analysis the dimensions of $\varphi$ and
$\overline{\varphi}$ must be equal to $\frac{d}{2}$. 
However, because of the symmetry breaking discussed above,
one has to allow for two distinct anomalous dimensions, $\eta$
and $\overline{\eta}$, of the fields $\psi$ and $\overline{\psi}$. 

We have worked  out the RG recursion relations for the couplings,
taking into account all one loop diagrams (they involve up to three vertices).
Some of these diagrams are shown in Fig.\,4.
For convenience we put 
\begin{equation}
X=\frac{g_1g_2}{4\lambda^2}K_4,\;\;Y=
\frac{u^2}{(\lambda+1)^2}K_4,\;\;Z=\frac{g_2u\mu}{\lambda^2(\lambda+1)^2}K_4
\end{equation}
With this notation we are led to the following recursion relations.
\begin{equation}\label{dg1}\begin{split}
\frac{\dd{g_1}}{\dd\ell}=&g_1\Big[\frac{\eps}{2}-\eta-\frac{\bar{\eta}}{2}-8
X+(3+\frac{1}{\lambda})Y-(5\lambda+3)Z\\&-\frac{(\lambda+1)^3}{4}\frac{Z
^2}{X}+\frac{(\lambda+1)^3}{4\lambda}\frac{YZ}{X}\Big]
\end{split}\end{equation}
\begin{equation}\label{dg2}
\frac{\dd
g_2}{\dd\ell}=g_2\Big[\frac{\eps}{2}-\frac{\eta}{2}-\bar{\eta}-8X+(3+\frac
{1}{\lambda})Y-2(2\lambda+1)Z\Big]
\end{equation}
\begin{equation}\label{du}
\frac{\dd
u}{\dd\ell}=u\Big[\frac{\eps}{2}-\frac{1}{2}(\eta+\bar{\eta})-4X+Y-
(2\lambda+1)Z\Big] 
\end{equation}
\begin{equation}\label{dlambda}
\frac{\dd
\lambda}{\dd\ell}=\lambda\Big[-\frac{1}{2}(\eta+\bar{\eta})-X+
\frac{1}{\lambda+1}Y+\frac{\lambda^2-4\lambda-1}{4(\lambda+1)}Z\Big]
\end{equation}
Upon requiring that the the coefficient of the
$\overline{\psi}\partial_t\psi$ term remain equal to unity we find the
supplementary equation
\begin{equation}
0=-\frac{1}{2}(\eta+\bar{\eta})-2X+Y-\frac{1}{2}(3\lambda+1)Z
\label{supplementary}
\end{equation}
\\ 
\noindent Finally, the coefficient $\lambda\sigma$ renormalizes according to
\begin{equation}\label{dsigma}\begin{split}
\frac{\dd\lambda\sigma}{\dd\ell}=&[2-\frac{1}{2}(\eta+\bar{\eta})]\lambda
\sigma+2\frac{g_1g_2}{2\lambda}\frac{K_4\Lambda^4}{\Lambda^2+\sigma}\\&-u^2
\frac{K_4\Lambda^4}{(\lambda+1)\Lambda^2+\lambda\sigma}\\&+\frac{ug_2\mu
}{\lambda}\frac{K_4\Lambda^6}{(\Lambda^2+\sigma)[(\lambda+1)\Lambda^2+
\lambda\sigma]}
\end{split}
\end{equation}
The preceding equations, together with the appropriate initial
conditions for $\ell=0$, determine the RG trajectories.
By combining Eqs.~(\ref{dsigma}) and (\ref{du}) we may 
check, before even solving the full set of equations, that to order $\eps$
\begin{equation}
y_\sigma=2-\frac{\bar{\eta}}{2}-4X^\star+Y^\star-(2\lambda+1)Z^\star=2-
\frac{\eps}{2}
\end{equation}
is independent of $\mu$. The existence of a Ward identity 
(see the appendix in \cite{KreeSchaubSchmittmann}) which is preserved
for $\mu\neq 0$ guarantees in fact that this result is true to all
orders in $\eps$.

\subsubsection{Equal diffusion constants, $\mu=0$}
\label{equal}
For $\mu=0$ the time reversal symmetry is restored, we must set
$\eta=\bar{\eta}$, and the RG equations Eqs.\,(\ref{dg1}) and (\ref{dg2}) for
$g_1$ and $g_2$ become identical.
The critical behavior is controlled by the fixed point
\begin{equation}
X^\star=\frac{\eps}{4},\;\;\;Y^\star=\frac{3\eps}{8},\;\;\;Z^\star=0,\;\;\;
\lambda=2,\;\;\;\eta=-\frac{\eps}{8}
\end{equation}
first found by Kree {\it et al.} \cite{KreeSchaubSchmittmann}.

\subsubsection{Unequal diffusion constants, $\mu\neq 0$}
\label{unequal}
The flow defined by Eqs.~(\ref{dg1}-\ref{dlambda}) 
and subject to condition (\ref{supplementary}) possesses a stable
fixed point at which $\lambda$ is the solution of 
\begin{equation}
2\lambda^3-9\lambda^2-6\lambda-1=0
\end{equation}
that is 
\begin{equation}
\lambda=[(2+\sqrt{3})^{1/3}+(2-\sqrt{3})^{1/3}-2]^{-1}=5.10664...
\end{equation}
and furthermore
\begin{equation}
X^\star=\frac{5\lambda^2+4\lambda+1}{8\lambda(3\lambda+1)}\eps=0.227\eps
\end{equation}
\begin{equation}
Y^\star=\frac{\lambda}{2(\lambda+1)}\eps=0.418\eps
\end{equation}
\begin{equation}
Z^\star=\frac{\lambda^2-2\lambda-1}{2\lambda(\lambda+1)(3\lambda+1)}\eps=0.0146\eps
\end{equation}
This leads to the critical exponents
\begin{equation}
\eta=0 \qquad \bar{\eta}=-\frac{\lambda}{3\lambda+1}\;\eps\,=-0.313\;\eps
\end{equation}
The result $\eta=0$ holds at every order in $\eps$. It follows that the
exponent $\beta=\nu(d+\eta)/2$ 
takes the mean field value
$\beta=1$. In checking the linear stability of this fixed point
in the subspace of $(g_1,g_2,u,\lambda)$ we found two
negative eigenvalues of multiplicity two. It is easy to see that this
fixed point governs the critical behavior of only the stationary state
with $\mu<0$. Indeed, the unrenormalized ($\ell=0$) value of $Z$
has the sign of $-\mu$; and from the definition of $Z$ combined with 
Eqs.\,(\ref{dg2}) and (\ref{du}) one deduces
that $Z$ cannot change sign under renormalization; so that 
the fixed point can be reached only if 
$Z(\ell=0)$ is positive just like $Z^\star$.

\subsection{A first-order transition for $\mu>0$ ?}
We shall now assume that the stationary state of the
system undergoes a first order transition for $\mu>0$. This assumption makes of
$\mu=0$ a tricritical point. We now develop a scaling analysis of the
stationary  state for $\mu$ of either sign. In this subsection the
exponents describing the $\mu<0$ fixed point will bear a minus index.\\
\noindent At the tricritical point $\mu=0$ characterized by $\eta,\,\nu$,
and $\beta$ the coupling constant $\mu$ is a relevant perturbation with
crossover exponent $y_{\mu}=-\eta/2>0$. For both $\sigma$ and $\mu$
small compared to microscopic momentum scales $\rho_B$ therefore has
the scaling form
\begin{equation}\label{rhoB1o}
\rho_B(\sigma,\mu)=(-\sigma)^{\beta}{\cal F}(\mu(-\sigma)^{-\nu y_{\mu}})
\end{equation}
The critical behavior for $\sigma\rightarrow 0$ at fixed $\mu<0$ is
described by the exponents $\eta_-\,,\nu$, and $\beta_-$ and we get
\begin{equation}
\rho_B(\infty)\simeq A (-\sigma)^{\beta_-}
\end{equation}
The amplitude $A$ is nonuniversal and is a singular function of $\mu$,
because the limits $\sigma\rightarrow 0$ and $\mu\rightarrow 0$ do not
commute. If  $\mu$ is small compared to microscopic scales but  large
compared to $(-\sigma)^{\nu y_\mu}$ we see from Eq.~(\ref{rhoB1o}) that 
\begin{equation}
A\sim (-\mu)^{(\beta-\beta_-)/(\nu y_\mu)}
\end{equation}
We use analogous arguments in the case $\mu>0$ assuming that the phase
transition is first order. A first order transition means that
$\rho_B(\infty)$ has a finite value in the limit $-\sigma\rightarrow 0$
at fixed $\mu>0$. Eq.~(\ref{rhoB1o}) now gives
\begin{equation}
\rho_B(\infty)\sim \mu^{1/\delta}
\end{equation} 
with $\delta=\nu y_\mu/\beta=-\eta/(d+\eta) > 0$.
Fig.\,6 shows a qualitative plot of the resulting behavior of the
stationary state density $\rho_B(\infty)$ as a function of $\rho$ in the
three regimes $\mu<0,\, \mu=0$, and $\mu>0$.

\subsection{Relaxation to the stationary state of the full action $S$}

\subsubsection{Non-Poissonian initial states: general remarks}
We wish to consider an initial particle distribution which, as
before, is the product on $i$ of identical single-site distributions
$p(m_i,n_i)$, where $m_i$ and $n_i$ count the $A$'s and $B$'s, 
respectively, on site $i$. The distribution $p$ is
arbitrary. It need not be Poissonian in either $m_i$ or $n_i$,
and need not factorize. 
The interest of studying such initial distributions is that all initial
distributions with short-range correlations reduce to them under
renormalization.  
The general distribution of this class is represented in
the action $S$ by the initial time terms
\begin{equation}
\int\,\text{d}^dx[-\rho_B^{(0)}\overline{\psi}(x,0)
+\Delta_\varphi\overline{\varphi}^2(x,0)
+\Delta_{\varphi\psi}\overline{\varphi}(x,0)\overline{\psi}(x,0)
+\Delta_\psi\overline{\psi}^2(x,0)]
\end{equation} 
plus terms of higher order in $\overline{\varphi}$ and
$\overline{\psi}$, which are  irrelevant under renormalization.
Here, setting $\ell_i=m_i+n_i$, we have
\begin{eqnarray}
\Delta_\psi&=&-\frac{1}{2\rho}[\langle\Delta n_i^2\rangle -
\langle n_i\rangle]\nonumber\\ 
\Delta_\varphi&=&-\frac{1}{2\rho}[\langle\Delta\ell_i^2\rangle -
\langle\ell_i\rangle]\nonumber\\ 
\Delta_{\varphi\psi}&=&-\frac{1}{\rho}[\langle\Delta n_i^2\rangle - \langle
n_i\rangle + \langle\Delta m_i \Delta n_i\rangle]
\label{Deltaphiphipsi}
\end{eqnarray}
If the $A$ and $B$ particles have uncorrelated Poisson distributions,
then $\Delta_\varphi=\Delta_{\varphi\psi}=\Delta_\psi=0$.
    
\subsubsection{Non-Poissonian initial $A$ distribution}
For the sake of simplicity we focus on the $\mu=0$ case. 
As discussed above, in an arbitrary initial state we have to deal with
the four parameters $\rho_B^{(0)},\, \Delta_\varphi,\,
\Delta_{\varphi\psi}$,\,  and
$\Delta_\psi$.
Also for simplicity, and because it is one of the interesting cases, we
shall consider here $\Delta_\psi=\Delta_{\varphi\psi}=0$.
Eq.\,(\ref{Deltaphiphipsi}) implies that this corresponds to having a Poisson
distribution of the $B$ particles, a {\it non}-Poissonian distribution
of the $A$ particles, and no correlations between the occupation numbers
of the two species. It then follows that
\begin{equation}
\Delta_\varphi=-\frac{1}{2\rho}[\langle\Delta m_i^2\rangle -
\langle m_i\rangle]
\label{Deltaphi}
\end{equation}
The $\Delta_\varphi$ term allows for contractions of the 
$\varphi$ between themselves according to 
\begin{equation}\label{phiphi}
\langle\varphi(k_1,t_1)\varphi(k_2,t_2)\rangle=-\Delta_\varphi
(2\pi)^d\delta^{(d)}(k_1+k_2)\Theta(t_1)\Theta(t_2)\text{e}^{-k_1^2(t_1+t_2)}
\end{equation}
There are no diagrams by which the $\Delta_\varphi$ term gets itself
normalized: it is strictly marginal. However, the new
contractions of Eq.\,(\ref{phiphi}) contribute to the renormalization of 
the $\rho_B^{(0)}$ term. The additional diagrams that appear are
shown in Fig.\,5. 
The  explicit expression of diagram Fig.\,5(b) reads
\begin{equation}\label{newcontrib}
u^2\Delta_{\varphi}\!\int\!\frac{d^dk}{(2\pi)^d}\!\int\! dt 
d\tau\psi(k,t)\overline{\psi}(-k,t+\tau)\Theta(t)\Theta(\tau)
\int_{q\in\Omega_{\Lambda}}\!\text{e}^{- 
2q^2 t-\lambda[(q-k)^2+\sigma]\tau-q^2\tau}
\end{equation}
One repeats the analysis performed in subsection \ref{relaxDP};
however, unlike diagram Fig.\,4(a) that appears in the directed
percolation case, diagram Fig.\,5(b) yields a renormalization of
$\rho_B^{(0)}$ that does not vanish at the fixed
point. The recursion relation for the $\rho_B^{(0)}$ vertex is
\begin{equation}\label{Deltapsi2}
\frac{\dd\rho_B^{(0)}}{\dd\ell}=\frac{d-\eta-\eta_0}{2}\rho_B^{(0)}
\end{equation}
where we have set
\begin{equation}\label{eta0}
\eta_0\equiv(\lambda+1)Y^\star \Delta_\varphi=\frac{9}{8}\Delta_\varphi\eps
\end{equation}
From Eqs.~(\ref{Deltapsi2}) and (\ref{eta0}) we conclude that the
scaling dimension of $\rho_B^{(0)}$ is
\begin{equation}
y_B=\frac{d}{2}-\frac{\eta}{2}-\frac{\eta_0}{2}
\end{equation}
This exponent is therefore nonuniversal. We may repeat the scaling
argument of section \ref{relaxDP2}, which allows us to conclude that the critical
initial slip exponent is 
\begin{eqnarray}
\theta'&=&-\frac{d+\eta}{2}+y_B=-\frac{\eta}{z}-\frac{\eta_0}{2z}
\nonumber\\
&=&\frac{2-9\Delta_\varphi}{32}\eps
\label{critslip}
\end{eqnarray} 
where in the last equality we have used that $z=2$.
We have therefore exhibited a nonuniversal exponent that characterizes
the short-time behavior of the density of the $B$ species.  
From expression (\ref{Deltaphi}), combined with the fact that at
criticality the density of $A$ particles differs negligibly from $\rho$,
one may deduce that 
\begin{equation}
\Delta_\varphi\leq \frac{1}{2}\rho
\label{Deltamin}
\end{equation}
so that $\theta'$ has the (nonuniversal) lower bound
$\theta^\prime_{\text{min}} = (4-9\rho)\eps/64$. 
The coupling $\Delta_\varphi$ attains the maximum value allowed by
Eq.\,(\ref{Deltamin}), and hence $\theta'$ its
minimum, when the $m_i$ have the probability law  
\begin{equation}
p_A(m_i)=(1-\rho)\,\delta_{m_i,0}+\rho\,\delta_{m_i,1}
\end{equation}
(where we take $\rho<1$). This means that the $A$ particles are 
randomly distributed but without
any multiple occupancy. Apparently, this is the environment giving the
slowest initial rise of the number of infected individuals. 

\subsubsection{Relaxation for $\mu<0$}
We shall now consider critical relaxation to the stationary
state for $\mu<0$, but limit ourselves to Poissonian initial
distributions: $\Delta_\psi=\Delta_\varphi=\Delta_{\varphi\psi}=0$.
The initial $B$ particle density renormalizes according to
\begin{equation}
\frac{{\text d}\rho_B^{(0)}}{{\text
d}\ell}=\frac{1}{2}(d-\bar{\eta})\rho_B^{(0)}
\end{equation}
After slightly generalizing the reasoning of subsection \ref{relaxDP2} we 
conclude that there is a nontrivial critical initial slip exponent 
\begin{equation}
\theta'=-\frac{\bar{\eta}}{2z}=\frac{\lambda}{4(3\lambda+1)}\eps=0.0782
\;\eps \qquad (\mu<0)
\end{equation}
To summarize, in the initial stage for $\mu<0$, $\rho_B(t)\sim
\rho_B^{(0)}t^{\theta'}$, then, over a time scale $\xi^{z}$, the
density decays as $\rho_B(t)\sim t^{-d/4}$. This last result holds at
every order in $\eps$.

\section{Conclusion}
We have adapted Wilson's renormalization scheme to the two-species
reaction--diffusion process  ~$A+B\rightarrow 2B,\,\, B\rightarrow A$. 
The stationary state of this process 
exhibits a second order phase transition at which the $B$'s
become extinct. 
A parameter $\mu=1-D_B/D_A$ controls the relative
strength of the diffusion constants $D_A$ and $D_B$ of the two species.
We have determined the critical behavior of the stationary state 
for $\mu<0$, that is, when the
the species subject to extinction ($B$) diffuses faster than the species
($A$) which it parasites on.
Special emphasis has been put on the relaxation from various initial 
states to the critical stationary state. 
As happens in magnetic systems, we found an initial
relaxation regime which scales with
an independent exponent $\theta'$, arising from renormalization of the
initial time terms in the action. The reaction--diffusion process has 
an order parameter (the $B$ particle density) that 
couples to a conserved quantity (the total density). 
We have shown that $\theta'$ varies continuously with the width of the
distribution of the conserved density. The same effect occurs in
magnetic systems with a continuous symmetry of the order parameter,
where the symmetry implies the existence of a conserved quantity
coupling to the order parameter \cite{JO}. 

We found that for positive $\mu$ the renormalization  
transformation has no stable fixed point, which signals the
absence of a second order phase transtion. We therefore discussed the
possibility of a first order transition; this scenario is plausible but
we lack decisive arguments in its favor. This unresolved point
deserves further attention.

A question of great interest concerns the low-dimensional behavior of
this reaction--diffusion 
system. The $\eps$ expansion is meaningful slightly below the upper
critical dimension $d_c=4$, and one may imagine that it still gives a
fairly good idea of what happens in $d=3$. However, the qualitative
picture provided by the $\eps$ expansion may break down below some
new critical dimension $d_c'$. 
Eq.\,(\ref{naivescaling}) shows that the (unrenormalized) action
contains terms, neglected in this work, that become relevant with
respect to the free (Gaussian) theory when $d=2$. It is therefore
possible that low
dimensions require separate consideration \cite{JEF}.
Even the seemingly simple one-dimensional version of this model 
exhibits nontrivial behavior \cite{MK} that remains to be elucidated.

\section*{Acknowledgments}
KO acknowledges useful discussions with J.L. Cardy and H.K. Janssen. His
work has been supported by the Deutsche Forschungsgemeinschaft (DFG).
HJH acknowledges stimulating interaction, during a visit to the
Universidade Federal do Rio Grande do Norte in Natal,
with Liacir S. Lucena and Joaquim E. de Freitas.
FvW and HJH thank Marumi Kado for his interest in this work and for 
the results of some one-dimensional simulations.
They also acknowledge a 
conversation with J.L. Lebowitz and H. Spohn on first-order transitions
in stationary states.


\newpage
\noindent FIGURE CAPTIONS\\

\noindent {\bf Figure 1.} Graphical conventions. The $\psi$ propagator
is shown in plain line, the $\varphi$ propagator appears as a dashed
line, and whenever it is non-zero, the $\langle\overline{\psi}\varphi\rangle$
correlator connects a plain leg to a dashed one. Also shown are the
vertices comprising the interaction term of the full action. A vertex
connecting legs by an empty circle $\circ$ is attached to $t=0$ and is
therefore not summed over time.\\

\noindent {\bf Figure 2.} Diagrams (a) and (b) renormalize
$\rho_B^{(0)}$ and $\Delta_\psi$, respectively. They are linear in
these quantities.\\

\noindent {\bf Figure 3.} Other one-loop order diagrams renormalizing
$\rho_B^{(0)}$ (diagram (a)) and $\Delta_\psi$ (diagrams (a) and (b)).\\

\noindent {\bf Figure 4.} Additional diagrams entering the
renormalization of $g_1$ for $\mu\neq 0$ with respect to the $\mu=0$ case.\\

\noindent {\bf Figure 5.} Diagrams (a) and (c) renormalize
$\Delta_{\psi}$. Diagram (b) renormalizes $\rho_B^{(0)}$ and $\Delta_\psi$. \\

\noindent {\bf Figure 6.} Stationary state density $\rho_B(\infty)$ of
the $B$ particles plotted qualitatively against
the total density $\rho$. The curve
for $\mu>0$ (dashed line) rests on the hypothesis of there being a first order
transition. The curve for $\mu=0$ (dotted line) was obtained in
Ref.[1]. The curve for $\mu<0$ (plain line) is a mean-field like straight line.

\newpage
\begin{figure}
\epsfxsize=\hsize
\epsfbox{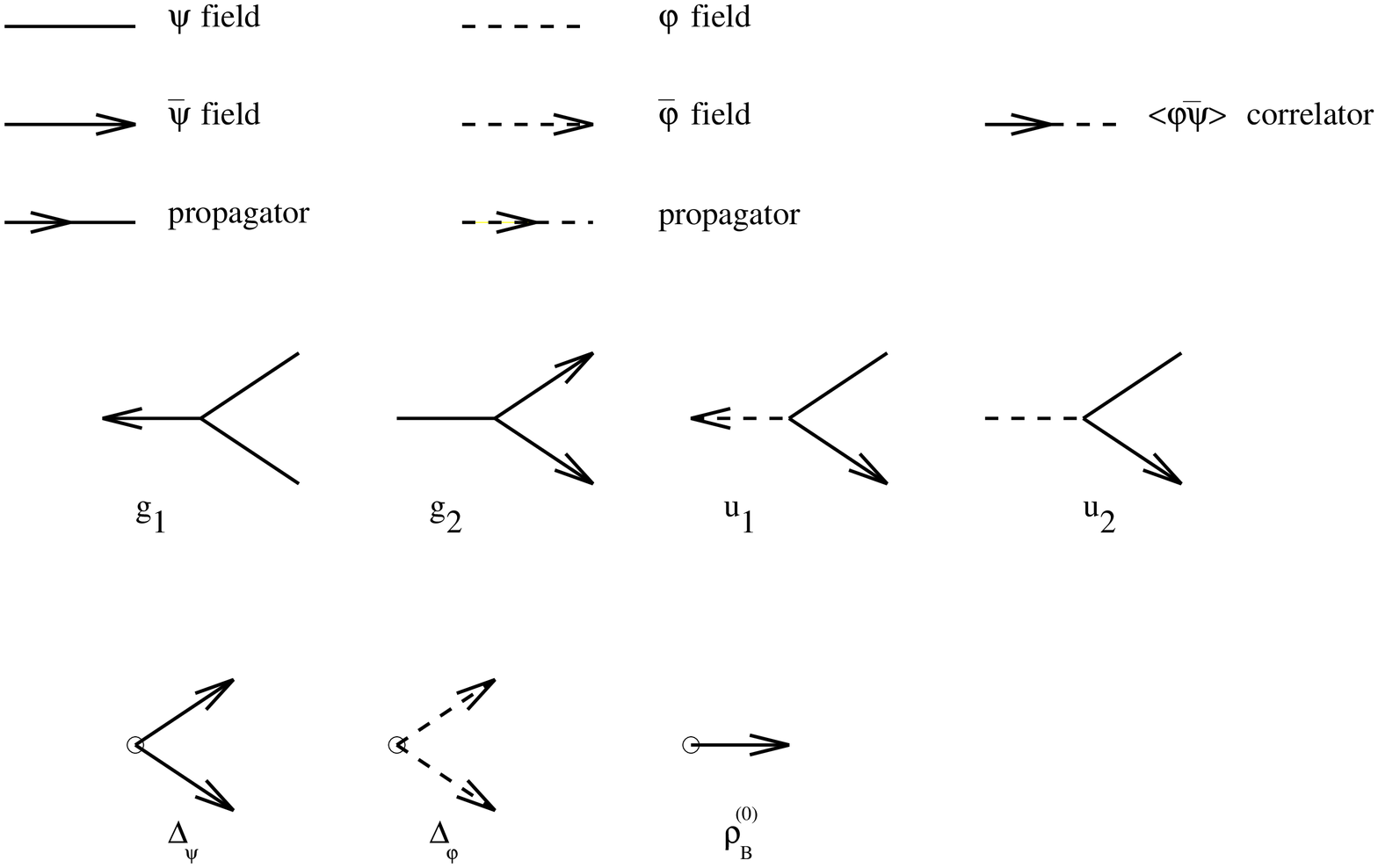}
\vskip 0.2in
\center{Figure 1}
\end{figure}

\begin{figure}
\epsfxsize=\hsize
\epsfbox{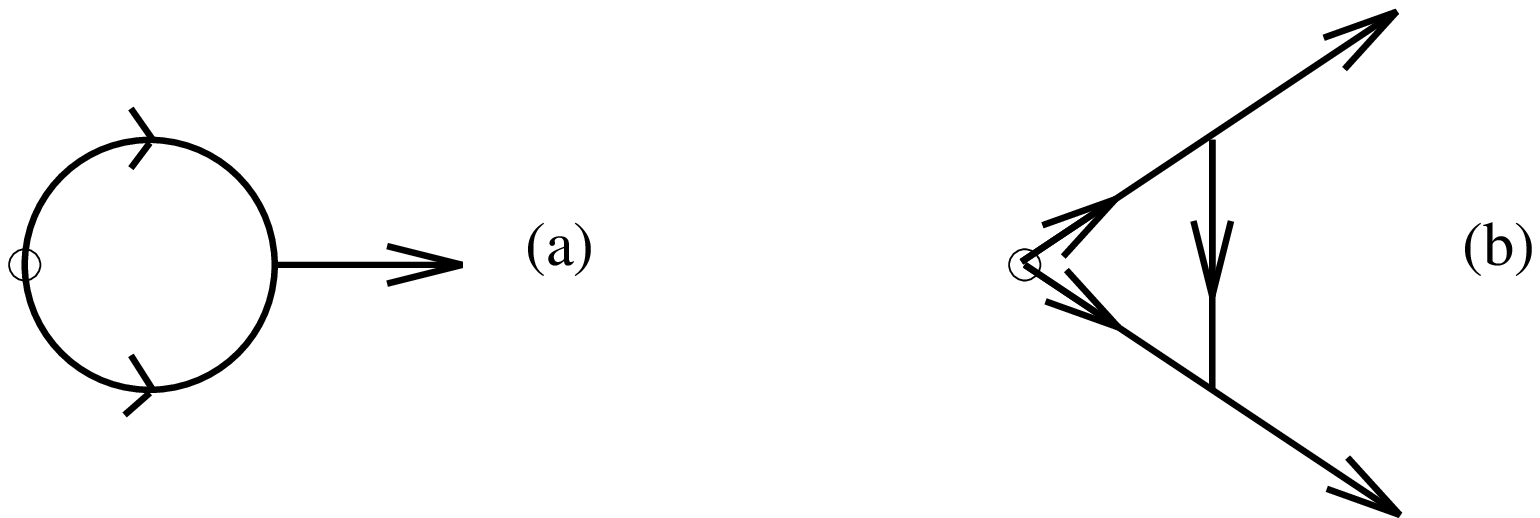}
\vskip 0.2in
\center{Figure 2}
\end{figure}

\begin{figure}
\epsfxsize=\hsize
\epsfbox{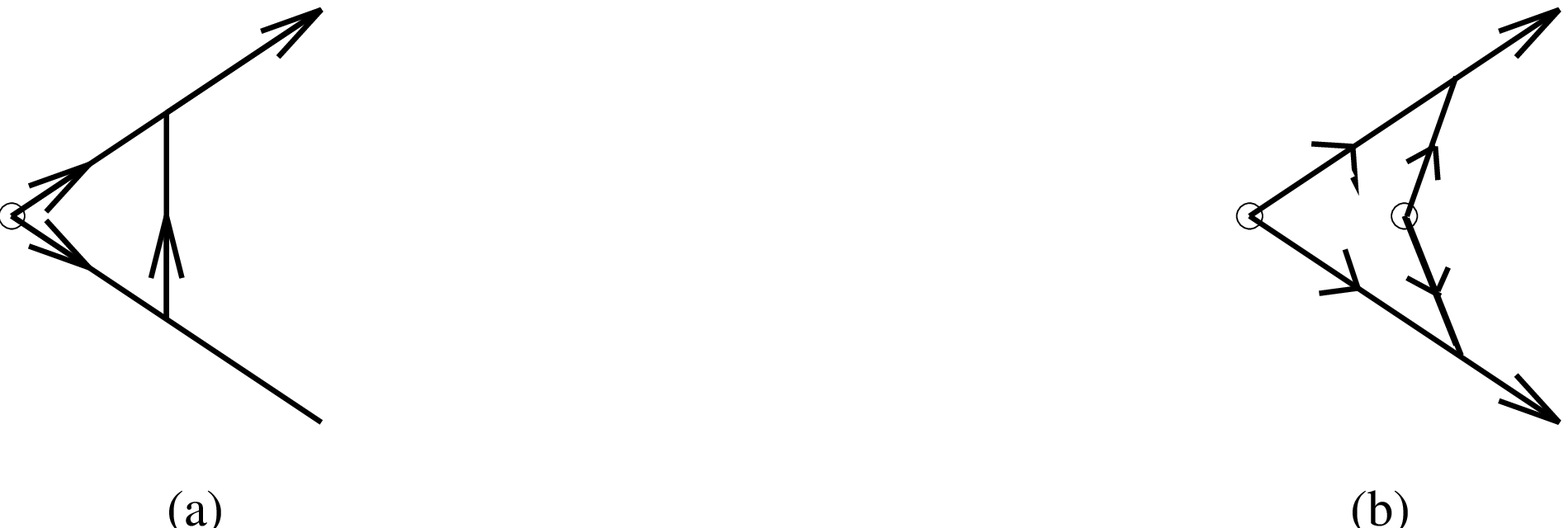}
\vskip 0.2in
\center{Figure 3}
\end{figure}

\begin{figure}
\epsfxsize=\hsize
\epsfbox{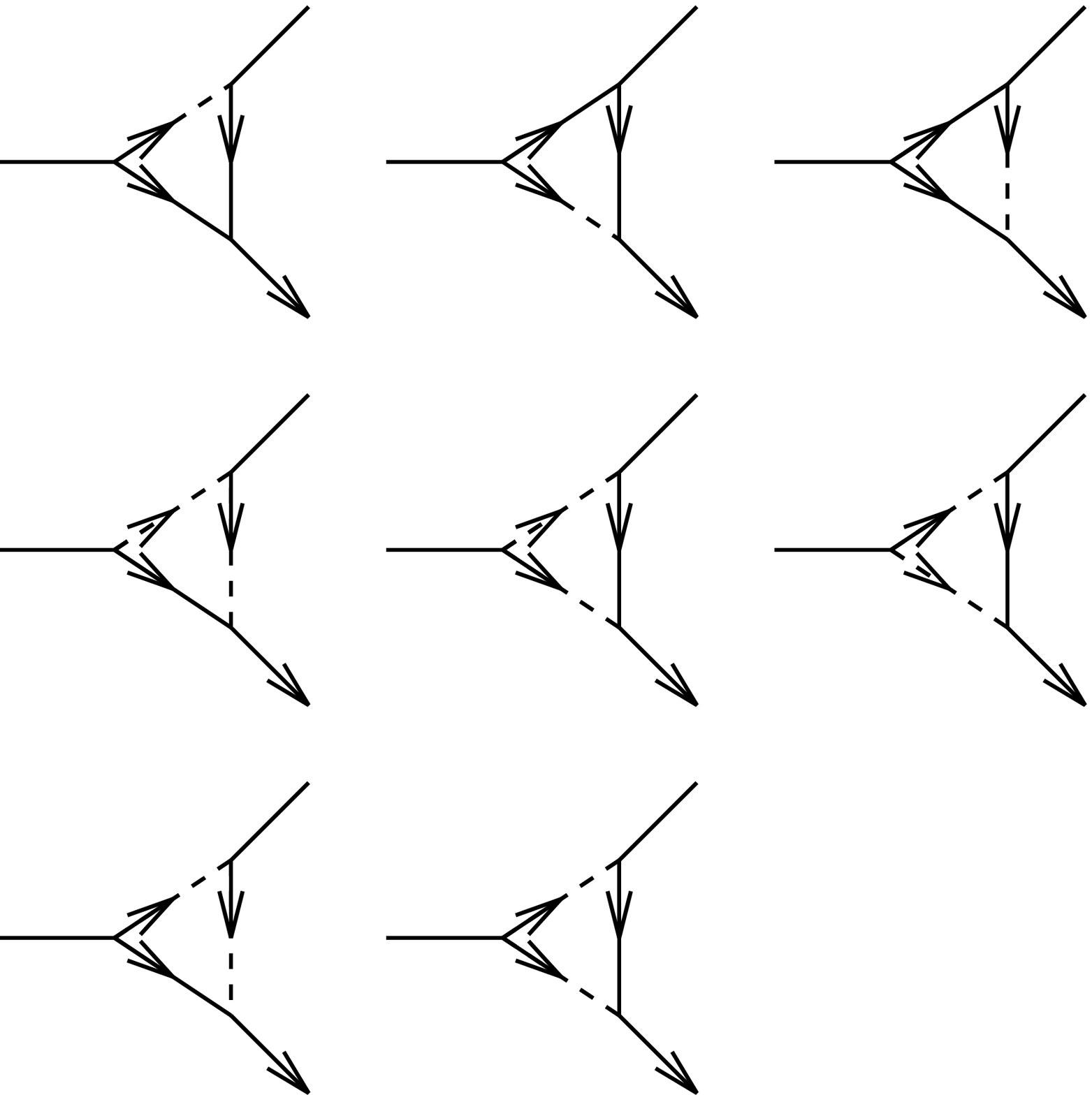}
\vskip 0.2in
\center{Figure 4}
\end{figure}

\begin{figure}
\epsfxsize=\hsize
\epsfbox{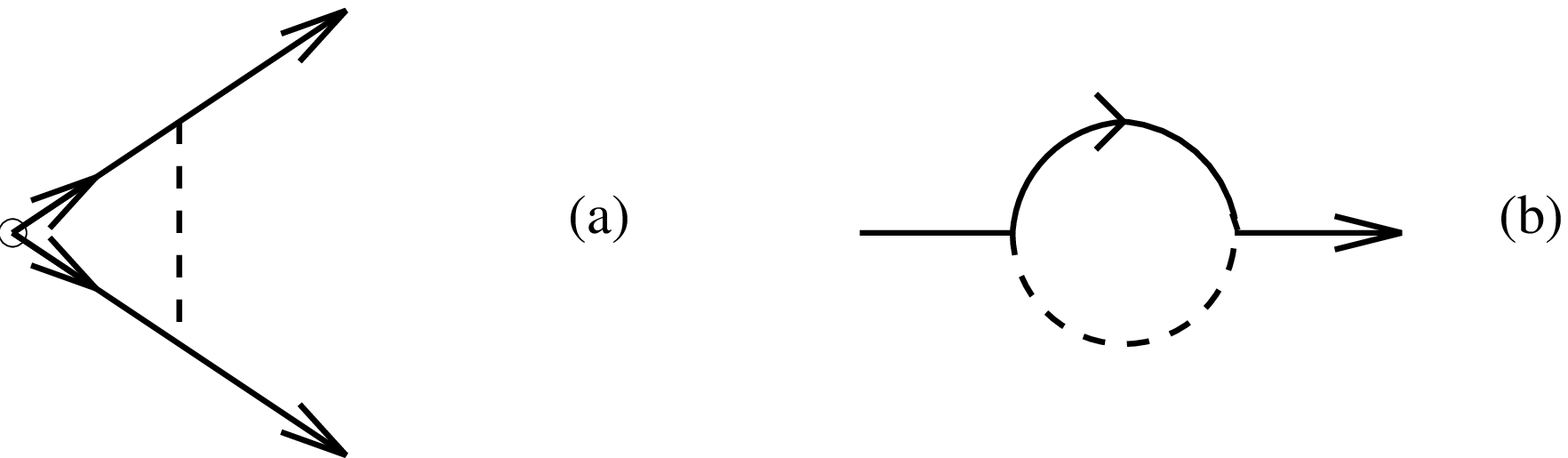}
\vskip 0.2in
\center{Figure 5}
\end{figure}

\begin{figure}
\epsfxsize=\hsize
\epsfbox{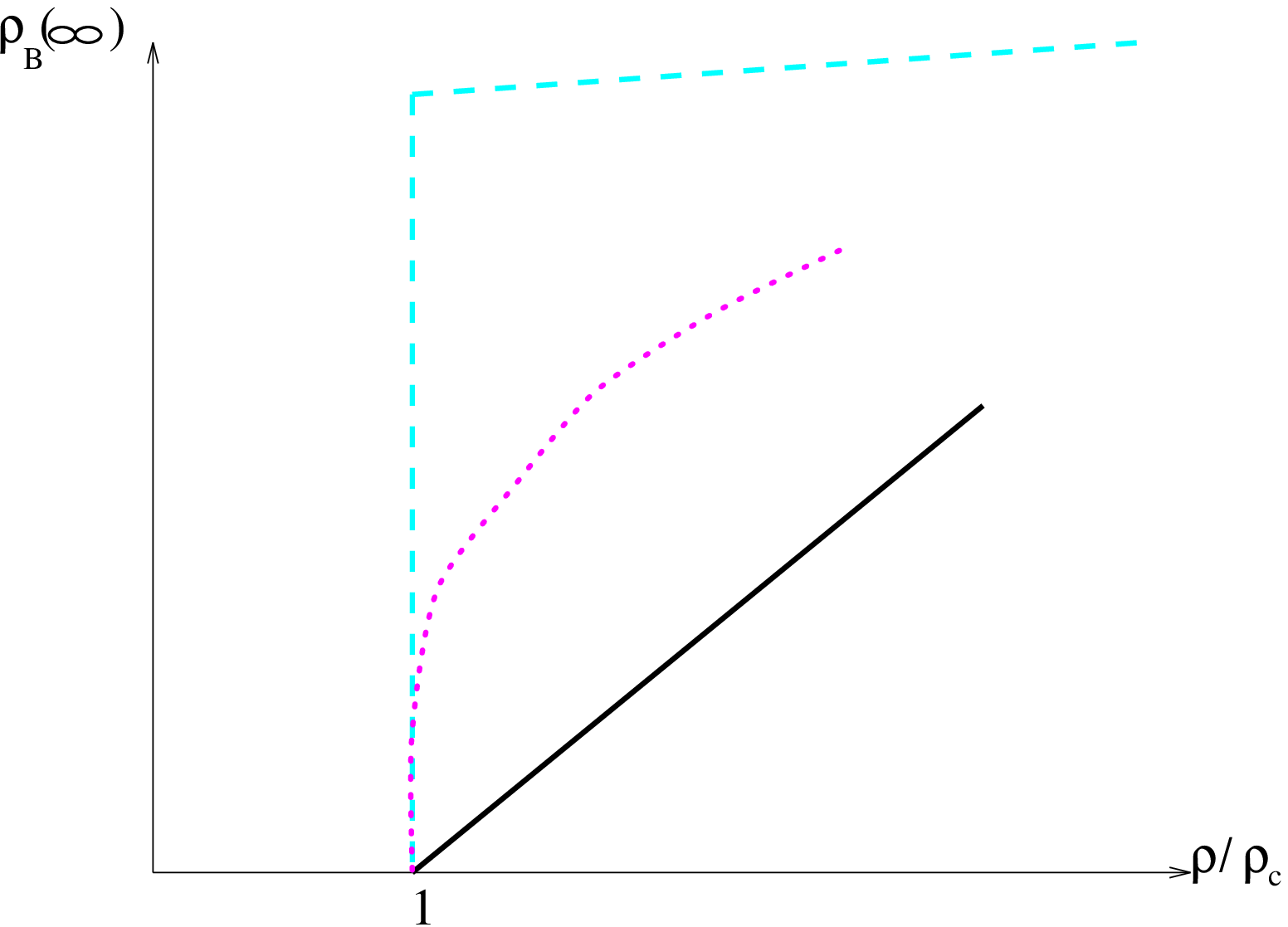}
\vskip 0.2in
\center{Figure 6}
\end{figure}


\newpage
\begin{thebibliography}{10}

\bibitem{KreeSchaubSchmittmann}
R. Kree, B. Schaub, and B. Schmittmann,
\newblock {\it Phys. Rev. A} {\bf 39} (1989) 2214.

\bibitem{JanssenSchaubSchmittmann}
H.~K. Janssen, B. Schaub and B. Schmittmann, 
\newblock {\it Z. Phys. B} {\bf 73} (1989) 539.

\bibitem{Janssenreview}
H.~K. Janssen, in 
\newblock {\it From Phase Transitions to Chaos}, edited by G.
Gy\"orgyi, I. Kondor, L. Sasv\'ari and T. T\'el (World Scientific, Singapore, 1992).

\bibitem{HohenbergHalperin}
P.~C. Hohenberg and B.~I. Halperin,
\newblock {\it Rev. Mod. Phys.} {\bf 49} (1979) 435.

\bibitem{Doi}
M. Doi,
\newblock {\it J. Phys. A} {\bf 9} (1976) 1465.

\bibitem{Peliti}
L. Peliti,
\newblock {\it J. Physique} {\bf 46} (1985) 1469.

\bibitem{LeeCardy}
B.~P. Lee and J. Cardy,
\newblock {\it J. Stat. Phys.} {\bf 80} (1995) 971.

\bibitem{CardySugar}
J.~L. Cardy and R.~L. Sugar, 
\newblock {\it J. Phys. A} {\bf 13} (1980) L423.

\bibitem{GrassbergerSundermeyer}
P. Grassberger and K. Sundermeyer,
\newblock {\it Phys. Lett. B} {\bf 77} (1978) 220.

\bibitem{GrassbergerdelaTorre}
P. Grassberger and A. de la Torre,
\newblock {\it Ann. Phys. NY} {\bf 122} (1979) 373.

\bibitem{J81}
H.~K. Janssen, 
\newblock {\it Z. Phys. B} {\bf 42} (1981) 151.

\bibitem{Schlogl}
F. Schl\"ogl,
\newblock {\it Z. Phys.} {\bf 253} (1972) 147.

\bibitem{Bronzan}
J.~B. Bronzan and J.~W. Dash,
\newblock {\it Phys. Rev. D} {\bf 10} (1974) 4208.

\bibitem{DiehlRitschel}
H.~W. Diehl and U. Ritschel,
\newblock {\it J. Stat. Phys.} {\bf 73} (1993) 1.

\bibitem{RitschelDiehl}
U. Ritschel and H.~W. Diehl,
\newblock {\it Phys. Rev. E} {\bf 51} (1995) 5392.

\bibitem{RC}
U. Ritschel and P. Czerner,
\newblock {\it Phys. Rev. E} {\bf 55} (1997) 3958.

\bibitem{JO}
K. Oerding and H.~K. Janssen, 
\newblock {\it J. Phys. A} {\bf 26} (1993) 5295.

\bibitem{JEF}
J.~E. de Freitas and L.~S. Lucena, unpublished simulations.

\bibitem{MK}
M. Kado, unpublished simulations.
\end{thebibliography}
\end{document}